\documentclass[conference]{IEEEtran}
\IEEEoverridecommandlockouts
\usepackage{cite}
\usepackage{amsmath,amssymb,amsfonts}
\usepackage{graphicx}
\usepackage{textcomp}
\usepackage[table]{xcolor}
\def\BibTeX{{\rm B\kern-.05em{\sc i\kern-.025em b}\kern-.08em
    T\kern-.1667em\lower.7ex\hbox{E}\kern-.125emX}}

\usepackage{booktabs}
\usepackage{color}

\usepackage{physics}
\usepackage{algorithm}
\usepackage{algpseudocode}
\makeatletter
\newcommand{\StatexIndent}[1][3]{%
  \setlength\@tempdima{\algorithmicindent}%
  \Statex\hskip\dimexpr#1\@tempdima\relax}
\makeatother

\usepackage{multirow}

\begin{document}

\title{Quantum {\it vs} classical genetic algorithms:\\ A numerical comparison shows faster convergence
\thanks{This work was supported by Spanish Government
Ram\'on y Cajal Grant RYC-2020-030503-I and the project grant PID2021-125823NA-I00 funded by MCIN/AEI/10.13039/501100011033 and by “ERDF A way of making Europe” and "ERDF Invest in your Future". Also by the QUANTEK project from ELKARTEK program (KK-2021/00070), as well as from QMiCS (820505) and OpenSuperQ (820363) of the EU Flagship on Quantum Technologies, and the EU FET-Open projects Quromorphic (828826) and EPIQUS (899368). R.I. acknowledges the support of the Basque Government Ph.D. grant PRE\_2021-1-0102.}
}

\author{\IEEEauthorblockN{Rub\'en Ibarrondo\IEEEauthorrefmark{1}\IEEEauthorrefmark{2},
Giancarlo Gatti\IEEEauthorrefmark{1}\IEEEauthorrefmark{2},
 and Mikel Sanz\IEEEauthorrefmark{1}\IEEEauthorrefmark{2}\IEEEauthorrefmark{3}\IEEEauthorrefmark{4}\IEEEauthorrefmark{5}}
\IEEEauthorblockA{\IEEEauthorrefmark{1}Department of Physical Chemistry, University of the Basque Country UPV/EHU, 48940 Leioa, Spain\\}\IEEEauthorblockA{\IEEEauthorrefmark{2}EHU Quantum Center, University of the Basque Country UPV/EHU, 48940 Leioa, Spain\\}\IEEEauthorblockA{\IEEEauthorrefmark{3}IKERBASQUE, Basque Foundation for Science, 48009 Bilbao, Spain\\}\IEEEauthorblockA{\IEEEauthorrefmark{4}Basque Center for Applied Mathematics (BCAM), Alameda de Mazarredo 14, 48009 Bilbao, Spain\\}\IEEEauthorblockA{\IEEEauthorrefmark{5}E-mail: mikel.sanz@ehu.eus}
}
\maketitle

\begin{abstract}
Genetic algorithms are heuristic optimization techniques inspired by Darwinian evolution. Quantum computation is a new computational paradigm which exploits quantum resources to speed up information processing tasks. Therefore, it is sensible to explore the potential enhancement in the performance of genetic algorithms by introducing quantum degrees of freedom. Along this line, a modular quantum genetic algorithm has recently been proposed, with individuals encoded in independent registers comprising exchangeable quantum subroutines \cite{Ibarrondo2022}, which leads to different variants. Here, we address the numerical benchmarking of these algorithms against classical genetic algorithms, a comparison missing from previous literature. To overcome the severe limitations of simulating quantum algorithms, our approach focuses on measuring the effect of quantum resources on the performance. In order to isolate the effect of the quantum resources on the performance, we selected the classical variants to resemble the fundamental characteristics of the quantum genetic algorithms.
Under these conditions, we encode an optimization problem in a two-qubit Hamiltonian and face the problem of finding its ground state. A numerical analysis based on a sample of 200 random cases shows that some quantum variants outperform all classical ones in convergence speed towards a near-optimal result. Additionally, we have considered a diagonal Hamiltonian and the Hamiltonian of the hydrogen molecule to complete the analysis with two relevant use-cases. If this advantage holds for larger systems, quantum genetic algorithms would provide a new tool to address optimization problems with quantum computers.
\end{abstract}

\begin{IEEEkeywords}
quantum computing, genetic algorithms, optimization
\end{IEEEkeywords}

\section{Introduction}
\label{sec:introduction}

Genetic Algorithms (GAs) are bioinspired algorithms with well-established performance in finding resilient solutions to complex optimization problems, such as problems with exponentially large search spaces and noisy optimization criteria \cite{Hornby2006, Holland1992, DeJong1993, Zebulum2001}. In these algorithms, every element of the search space can be potentially represented by an individual and the selection towards the optimal solution is performed by Darwinian-like evolution. The set of individuals is known as population, their codification is commonly known as chromosomes, and their performance is ranked by a fitness function. Although there is no formal definition of GA which univocally distinguishes them from other evolutionary algorithms, there is a general consensus about the presence of the following four characteristic elements: population-based search through joint evolution of a set of individuals, a selection of some of them according to their performance, a crossover operation to breed new individuals, and a mutation operation which randomly modifies them \cite{Melanie1996}.

During the past decades, the merge of GAs and quantum computation has been a source of new heuristic optimization methods \cite{Sofge2008, Roy2014, Lahoz-Beltra2016}. Encouraged by the principles of quantum mechanics, most of the effort has been oriented to quantum inspired GAs, which mimic quantum superposition and qubit rotations to engineer new varieties of classical evolutionary algorithms \cite{Narayanan1996, Han2000, Han2001, Han2002, Yang2004, Wang2005, Yingchareonthawornchai2012, Roy2014}.  
On the other hand, fully quantum approaches potentially achieving quantum speed-up have only attained partial success in the inclusion of the aforementioned characteristic elements \cite{Rylander2001, Udrescu2006, Malossini2008, Saitoh2014}. In 2001, Rylander et al. proposed in Ref. \cite{Rylander2001} a quantum genetic algorithm introducing the concepts of chromosome-register and fitness-register. Although it was claimed that quantum superposition could provide an increased searching power, this conclusion has been considered unsupported due to the lack of heuristic or analytic evidence \cite{Sofge2008}. In 2006, Udrescu et al. proposed in Ref. \cite{Udrescu2006} an algorithm based on quantum searching and inspired by evolutionary computation. In 2008, Malossini et al. \cite{Malossini2008} proposed an optimization algorithm that also used quantum search to enhance the selection subroutine in a GA. These two proposals were extended by Saitoh et al.~\cite{Saitoh2014} including quantum crossover and mutation in the algorithm. Nevertheless, as the whole population is encoded in the superposition of a single register, the selection procedure is implemented by projective measurements. Therefore, only one individual is selected in each generation, consequently reducing the intrinsic exploration capacity with respect to GAs.

In Ref.~\cite{Ibarrondo2022}, the authors presented a quantum genetic algorithm (QGA) that comprised all the fundamental elements of GAs as fully quantum operations by means of encoding individuals in the state of qubit-registers. The algorithm is composed of modular quantum subroutines inspired by classical GAs: selection, crossover, and mutation. This modular structure allowed the authors to study two QGA variants arising from two paradigmatic approximated quantum cloning machines employed to breed new individuals. Although they presented figures of merit to evaluate the performance of these algorithms and compare QGA variants, they did not address a benchmark against classical algorithms. This benchmark is important to determine the viability of QGAs as optimization methods. Testing this requires the design of a careful comparison which takes the cost of quantum simulation into account.

In this article, we perform the numerical comparison of the QGA with classical genetic algorithms. In Section~\ref{sec:description_gas}, we state our premises to select proper classical counterparts to compare with the QGA and describe the chosen algorithms, together with a brief review of the QGA. In Section~\ref{sec:comparison}, we first lay the foundations of the methodology used for the comparison, specifying the chosen parameters and problem-cases. Afterwards, we numerically compare the performance of each algorithm. Finally, we discuss some conclusions in Section~\ref{sec:conclusion}.

\section{Description of the compared algorithms}
\label{sec:description_gas}

Here, we summarize the fundamental structure of the QGA proposed in Ref.~\cite{Ibarrondo2022} and introduced the classical GAs considered for the comparison. The considered optimization problem is encoded into a problem  Hamiltonian $H_P$ which describes the energy of a $c$-qubit system, and whose ground state represents the optimal solution. Hence, the optimization algorithms search the Hilbert space $\mathcal{H}$ associated with a $c$-qubit system minimizing the energy of their individuals encoded as quantum states. Regarding the classical GAs, we make the choice to study two methods for the encoding of the individuals, namely as complex column vectors or as bit-strings. The former is the most general encoding because it can be used to represent any possible state in $\mathcal{H}$, whereas bit-strings can only be used to represent a discrete number of states, which is useful if the eigenbasis of the Hamiltonian is known and we can associate each state to a bit-string. Although the latter has restricted applicability, it has been included for the sake of completeness. More sophisticated encoding strategies can be proposed to suit a particular problem if it possesses symmetries or constraints, however, we have pursued simplicity to be able to study the effect of quantum resources. For the same reason, we have avoided too many random variables and opted for a unique selection method on the classical GAs. This choice allows us to identify and highlight the fundamental differences between QGAs and classical GAs in Section~\ref{sec:comparison}.

\begin{table}
\centering
\caption{Summary of the acronyms for the compared GAs. There are four quantum variants, four classical variants with complex-vector representation, and a single classical variant with bit-string representation. Variants are classified in terms of the cloning or crossover method, and the mutation method described in Section~\ref{sec:description_gas}.}
\renewcommand{\arraystretch}{1.7}
\begin{tabular}{llcc}
                                          &               & \cellcolor{black!10} No mutation  & \cellcolor{black!10} With mutation \\ 
\multirow{2}{4em}{\textbf{Quantum}}                & \cellcolor{black!10} BCQO          & QGAbnm       & QGAbwm        \\
                                          & \cellcolor{black!10} UQCM          & QGAunm       & QGAuwm      \\ \addlinespace[0.5ex] \hline
\addlinespace[0.5ex]
                                          &               &  \cellcolor{black!10} Mutation (i) & \cellcolor{black!10} Mutation (ii) \\
\multirow{2}{4em}{\textbf{Classical complex-vector}} & \cellcolor{black!10} Crossover (a) & CGAai        & CGAaii        \\
                                          & \cellcolor{black!10} Crossover (b) & CGAbi        & CGAbii        \\ \addlinespace[0.5ex] \hline 
\addlinespace[2ex]
\multirow{1}{4em}{\textbf{Classical bit-string}}                      &               & \multicolumn{2}{c}{BGA}  \\
\end{tabular}
\label{tab:compared_algs}
\end{table}

\subsection{Quantum genetic algorithm}
\label{sec:description_qga}

In Ref.~\cite{Ibarrondo2022} the authors proposed a Quantum Genetic Algorithm (QGA) comprising all the fundamental elements of GAs. This work is built upon the representation of individuals by states of independent quantum registers and considered the minimization of a problem Hamiltonian $H_P$. The population is encoded into the joint state of $n$ registers which comprise $c$ qubits each for a total of $nc$ qubits. This allows for the exploration of the search space $\mathcal{H}$ through Darwinian-like evolution. Here we provide a schematic overview of the QGA in Algorithm~\ref{alg:qga} and remark on the main features of the quantum subroutines.

\begin{algorithm}[htb]
\caption{Quantum genetic algorithm}\label{alg:qga}
\begin{algorithmic}[1]
\State $n \gets number\_of\_registers$ \Comment{assumed divisible by four}
\State $c \gets number\_of\_qubits\_per\_register$ \Comment{assumed even}
\State Initialize with a random state
\Repeat
	\State \textbf{sort} registers $1$ to $n$
	\State \textbf{reset} registers $n/2$ to $n$
	\For{ $r=1,2,\dots, n/2$}
		\State \textbf{pseudo-clone} register $r$ to register $n/2+r$.
	\EndFor
	\For{ $i=1,2,\dots, n/4$}
		\State \textbf{swap} the last $c/2$ qubits of register $n/2 + 2 i - 1$
	    \StatexIndent[2] with the last $c/2$ qubits of register $n/2 + 2 i$.
	\EndFor
	\State \textbf{mutate} each qubit with probability $p_m$
 \Until{ending criteria is met $\lor$ $G$ generations} 
\end{algorithmic}
\end{algorithm}

The core of the quantum selection subroutine is a quantum sorting network \cite{Berry2018}. This protocol uses an ancillary work-space to store the pairwise comparisons between the registers while it performs the required permutations to sort the population. Afterwards, individuals in the lowest half of the population are discarded. Instead of performing the pairwise comparisons in terms of the average energy, the states are compared in the eigenbasis $\qty{\ket{u_i}}_{i=1}^{2^c}$ of $H_P$ with ordered energies $\lambda_{i}\leq \lambda_{i+1}$. For instance, let $\sum_{i,j=1}^{2^c} b_{i,j} \ket{u_i}_1\ket{u_j}_2\ket{0}_a$ be the initial state of registers $1$, $2$ and an ancilla, then the sorting network transforms it into
\begin{equation}
\sum_{i,j=1}^{2^c} b_{i,j}
\begin{cases}
\ket{u_i}_1\ket{u_j}_2\ket{0}, & \mathrm{if}\; \lambda_i \leq \lambda_j, \\
\ket{u_j}_1\ket{u_i}_2\ket{1}, & \mathrm{if}\; \lambda_j < \lambda_i.
\end{cases}
\end{equation}
Note that the ancillary work-space is required so that the operation is reversible, and that discarding its state and part of the population implies that, in general, the final state is a mixture of sorted populations.

Quantum crossover implements the replication of the selected individuals to populate the discarded registers. The first step requires initializing the registers that have been discarded and consequently applying an approximate quantum cloning machine (QCM) \cite{Scarani2005}. Finally, swap operations are performed between half of the qubits of the lower individuals in order to combine their states. As there are different QCMs this step can introduce different evolutionary patterns depending on our choice. We consider two QCMs, namely the biomimetic cloning of quantum observable (BCQO) \cite{ARodriguez2014} and the Bu\v zek-Hillery universal quantum cloning machine (UQCM) \cite{Buzek1996}. The BCQO perfectly clones the statistics related to an observable, but its fidelity with respect to ideal cloning depends on the input state. In contrast, the UQCM approximately clones every state with identical fidelity, but it does not clone any state perfectly.

Finally, the quantum mutation subroutine consists in applying single qubit Pauli gates in each qubit of the population with a small probability $p$. However, it has also been proposed to suppress this subroutine in the QGA, due to its minor impact on the performance for many problem instances. Therefore, we have considered BCQO with mutation (QGAbwm) and without mutation (QGAbnm), and UQCM with mutation (QGAuwm) and without mutation (QGAunm).

\subsection{Classical genetic algorithms}
\label{sec:description_cga}

For the classical GAs, we have employed two different encodings for the individuals, namely, bit strings and complex vectors. The former is the most direct analog to the QGA, but its scope of tractable problems is constrained to diagonal Hamiltonians.

\paragraph{Bit-string GA (BGA)} The $n$ individuals are represented by classical $c$-bit strings which have definite energy with respect to $H_P$, i.e. individuals represent eigenstates. Selection is performed by sorting the individuals according to their energy and only preserving the $n/2$ ones with the lowest values. Afterwards, they are duplicated and the bit-strings of the copies are combined by swap operations. Precisely, the copies are grouped by pairs according to their energies, then they preserve the first $c/2$ bits and concatenate the last $c/2$ bits of the other. Finally, the mutation is introduced by bitwise \textsc{not} operation with a probability $p$.

\paragraph{Complex-vector GAs (CGAs)} The $n$ individuals are encoded in complex column vectors of size $2^c$, so they represent an arbitrary state of $\mathcal{H}$. Selection is performed by sorting the individuals according to their average energy and preserving the half with the lowest energy. Afterwards, the latter are duplicated and their copies combined in two different manners:
\begin{itemize}
\item[a.] Performing a linear combination of them, i.e. let $v_1$ and $v_2$ be the column associated to the copied individuals, then the combinations are $2v_1+v_2$ and $v_1+2v_2$ normalized.
\item[b.] Interchanging half of the coefficients of the column vectors, so that they preserve the first $2^c/2$ coefficients and take the remaining ones from the other.
\end{itemize}
Finally, we also propose two mutation methods:
\begin{itemize}
	\item[i.] Adding small vectors whose coefficients are computed by sampling a normal distribution with a small standard deviation $\sigma$ and multiplying it with a Bernoulli variable with the probability of one equal to $q$, which facilitates that there are unchanged coefficients.
	\item[ii.] Randomly applying single-qubit Pauli gates with probability $p$.
\end{itemize}
Therefore, we consider four complex vector GAs noted CGAai, CGAaii, CGAbi, and CGAbii indicating the crossover and mutation method used.

\section{Comparison of the QGA and the classical GAs}
\label{sec:comparison}

\begin{figure*}[!htb]
\centering
\includegraphics[scale=1]{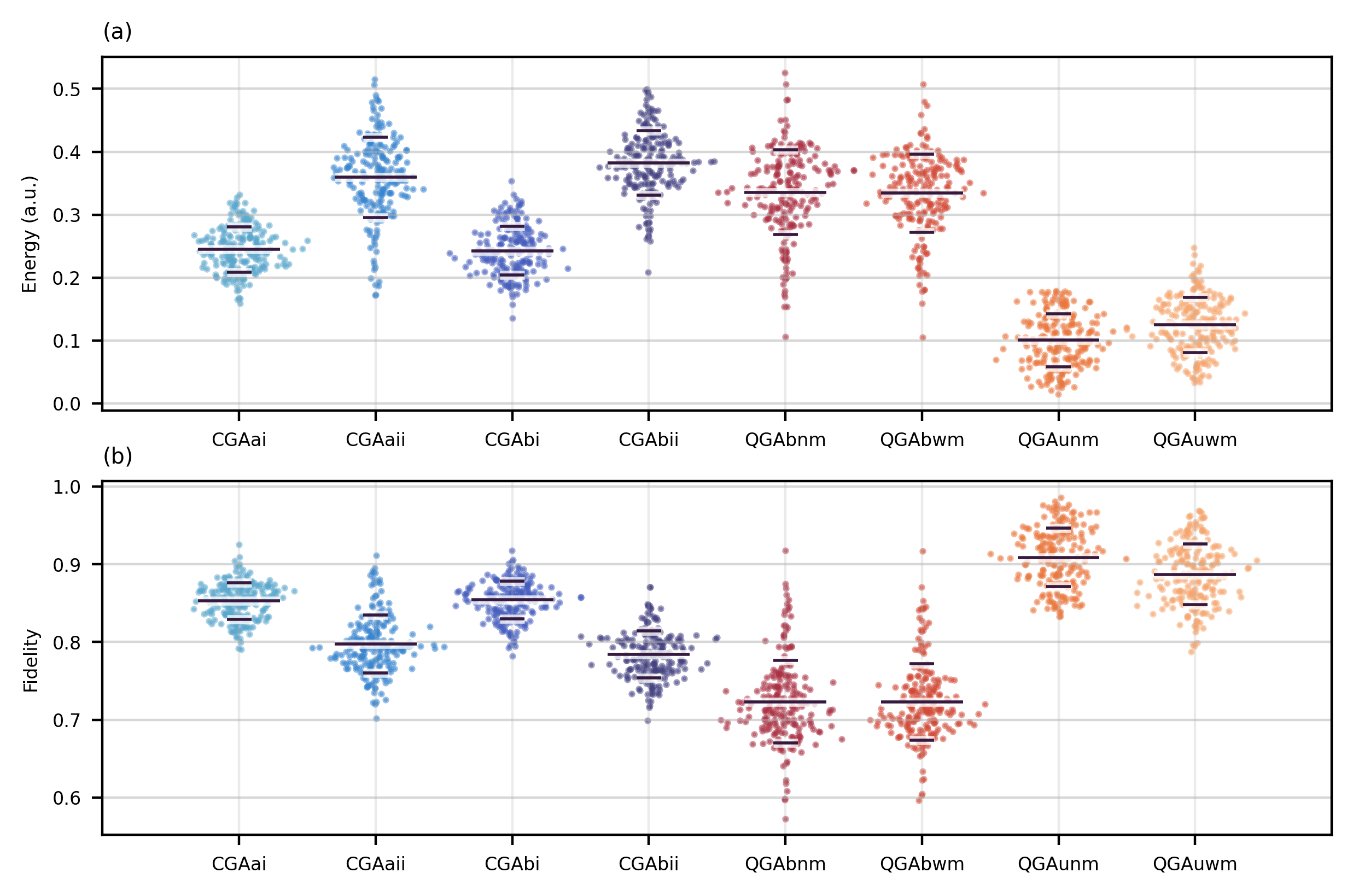}
\caption{(a) Mean energy of the lowest-energy individual and (b) fidelity between this individual and the desired Hamiltonian ground state after $10$ generations. The fidelity is the standard measure of similarity between quantum states and is computed by $\expval{\rho}{u_1}$ for arbitrary state $\rho$ and the ground state $\ket{u_1}$. We considered four CGAs and four QGAs, selected as described in Section~\ref{sec:description_gas}, and we applied each of them to a sample of randomly generated problem Hamiltonians. The vertical position of each point is the $0.90\%$ quantile value for CGAs and the average value for QGAs from several initial populations for each $H_P$, and the horizontal position is jittered to reduce the visual overlap. The horizontal black lines depict the average and standard deviation for each GA, the numerical values are summarized in Table~\ref{tab:stats_data}. Noticeably, the best quantum variant is UQCM without mutation and has a mean and standard deviation equal to $0.91(0.04)$, greater than the best classical variants which have $0.85(0.02)$ and $0.85(0.02)$, for CGAai and CGAbi respectively.}
\label{fig:qga_vs_cga-top10}
\label{fig:qga_vs_cga_energy-top10}
\end{figure*}

QGAs and CGAs essentially differ in the state of the population, and both differ from BGA in their effective search space. Individuals are arbitrary quantum states in the former cases, but only the quantum counterpart admits a superposition at a population level, i.e. only QGAs enable entanglement between individuals. This restriction of the classical algorithms is imposed by the fact that the amount of resources scales with $2^n$ if entanglement between $n$ quantum registers is included. However, this fundamental difference entails features that can be regarded both as positive and negative for each approach.

On the one hand, the ability to operate with entangled individuals enables the quantum sorting subroutine to compare in the eigenbasis of the problem Hamiltonian. Whereas, separability-preserving classical sorting has to be performed with other criteria, such as mean energy. Therefore, the selection process cannot specifically pick the components with lower energy but is forced to select the whole state. Thus, we expect a superior ability to distillate successful features of the individuals in the quantum selection subroutine.

On the other hand, a fully quantum representation of the population comes along with constraints imposed by fundamental quantum mechanics. Indeed, it is impossible to perfectly clone arbitrary quantum states and to reset arbitrary states in a reversible manner, as stated by the no-cloning \cite{Wootters1982} and no-deleting theorems \cite{Pati2000}. Consequently, CGAs have the advantage of preserving all the features of selected individuals, whereas QGAs inevitably lose a part. Conversely, CGAs can erase discarded individuals without affecting the selected ones, but QGAs inevitably alter them in this process, due to quantum entanglement.

These differences result in the fundamental notion that QGAs can be described by quantum channels, i.e. the most general input-output transformations that quantum states can undergo. As shown in Ref.~\cite{Ibarrondo2022}, QGAs converge exponentially towards stationary states, i.e fixed-points of the quantum channel, which are near to the optimum solution. Therefore, we expect that QGAs obtain low energy states faster than classical counterparts, but also that CGAs should eventually be able to overcome the stationary state of the QGAs by investing a larger amount of generations. In Section~\ref{sec:numerical_performance}, we discuss in detail the effect of these features on numerical performance.

Beyond this comparison, an exhaustive analysis of the demanded resources and time-cost of the algorithms would be desirable in the future. In general, we expect an advantage in terms of the representation of the individuals, as representing quantum states in classical computers requires in general an exponential amount of resources\footnote{Of course, this advantage in the representation holds only if we do not need complete knowledge about final state, which would require full tomography and an exponential number of repetitions of the QGA.}, and a similar dependence of the time-cost of the subroutines with the size of the individuals. However, that comparison would have to take into account the precise quantum hardware, its allowed connectivity, and its set of fundamental operations, in order to support these claims.

\subsection{Numerical performance of the algorithms}
\label{sec:numerical_performance}

In this Section, we compare the performance of the classical GAs with the performance of the QGA addressing different problem instances. The size of the populations for all GAs was $n=4$ individuals and the search space was the Hilbert space associated with $c=2$ qubits, i.e. we used $2$ qubits per individual in the QGAs, four-dimensional complex vectors for CGAs and two-bit string for the BGA. The figures of merit for this benchmark are the quantum fidelity between the best individual and the desired Hamiltonian ground state, and the expected energy. The quantum fidelity is a standard measure of similarity between quantum states that ranges from $0$ to $1$ and for a density matrix $\rho$ representing the state of the individual and the ground state of the Hamiltonian $\ket{u_1}$ is given by
\begin{equation}
\label{eq:fidelity}
F(\rho, \ket{u_1}) = \expval{\rho}{u_1}.
\end{equation}
The energy of a state $\rho$ with respect to the problem Hamiltonian $H_P$ is given by
\begin{equation}
\label{eq:energy}
\expval{E}_{\rho} = \tr[H_P \rho].
\end{equation}
Although the average energy is a standardized figure of merit for numerically rating the performance, the fidelity is a stronger metric for theoretical analysis, as it also ensures the similarity of the outcome for any observable measurement.
The first benchmark consists of the randomly generated sample of 200 problem Hamiltonians used in Ref.~\cite{Ibarrondo2022}, whose elements differ in eigenbasis but have equal spectrum given by integer numbers in arbitrary units (denoted a.u.). Additionally, we included the
molecular electronic Hamiltonian of the hydrogen molecule in the Bravyi-Kitaev Pauli representation $H_{H_2}$ given in atomic units \cite{Seely2012}, i.e. in units of Hartree energy denoted E$_h$, and a diagonal Hamiltonian $H_C$ given in arbitrary units:
\begin{eqnarray}
\label{eq:hh2}
H_{H_2} & = & 
\mqty(\dmat{0.469, 0.216 & 0.181\\ 0.181 & -1.361, 0.676}),
\\
\label{eq:hc}
H_C & = & 
\mqty(\dmat{0, 1, 2, 3}).
\end{eqnarray}
The $H_{H_2}$ provides a physically meaningful Hamiltonian whose ground state energy is $-1.382$ E$_h$, whereas the $H_{C}$ allows us to compare with the BGA and has zero ground state energy. Let us note that these Hamiltonians have the particularity that their eigenbasis is nearly (or exactly) the canonical basis, thus we observe some features that differ from the insights derived from the sample of Hamiltonians.

Simulations were performed with matrix computations in Python-NumPy. The QGAs were let to evolve for 10 generations for the sample of Hamiltonians repeating the process for 10 different initial populations, whereas for classical GAs 10 generations and 100 initial populations were used. For the case of $H_{H_2}$ and $H_C$, we employed 50 generations and 50 different initial populations for all the GAs. We used a mutation probability $p=q=\frac{1}{24}$ for QGAs and classical GAs, and a standard deviation $\sigma = 0.228$ for the first mutation method of CGAs.

\subsubsection{Results for the sample of Hamiltonians}

As the quantum algorithms have a single fixed point for this set of Hamiltonians, i.e. any initial state converges to the same final population, both the final fidelity and the final energy are accurately represented by their mean values. However, the final population for the CGAs strongly varies depending on the initial population and the mutation pattern, therefore we used the $0.90$ quantile of the results obtained for different initial populations, i.e. the fidelity value that is only surpassed by $10\%$ of the initial populations for the same problem Hamiltonian.

\begin{table}
\centering
\caption{Mean and standard deviation of the final fidelity and energy of each GA applied to the random sample of 200 Hamiltonians.}
\renewcommand{\arraystretch}{1.1}
\begin{tabular}{ccc}
			&  Fidelity & Energy (a.u.) \\[1ex] \hline
\rule{0pt}{3ex}
CGAai       & $0.85 (0.02)$ & $0.66 (0.04)$ \\
CGAaii      & $0.80 (0.04)$ & $0.72 (0.04)$ \\
CGAbi       & $0.85 (0.02)$ & $0.63 (0.04)$ \\
CGAbii      & $0.78 (0.03)$ & $0.73 (0.04)$ \\
QGAbnm      & $0.72 (0.05)$ & $0.34 (0.07)$ \\
QGAbwm      & $0.72 (0.05)$ & $0.33 (0.06)$ \\
QGAunm      & $0.91 (0.04)$ & $0.10 (0.04)$ \\
QGAuwm      & $0.89 (0.04)$ & $0.12 (0.04)$ \\[1ex]\hline
\end{tabular}
\label{tab:stats_data}
\end{table}

Figure~\ref{fig:qga_vs_cga_energy-top10} shows the values obtained for the final (a) mean energy and (b) fidelity with each GA. Each point represents the value obtained for a different Hamiltonian, in the case of the CGAs the value of the vertical axis is the (a) $10\%$ quantile of the energy and (b) $90\%$ quantile of the fidelity in the distribution of the final state obtained from different initial populations, whereas in the case of the QGAs it is the average value. The black horizontal lines represent the average and the standard deviation of the values for different problem instances, the precise numbers are summarized in Table~\ref{tab:stats_data}. According to this figure\footnote{We used a Wilcoxon signed-rank test to analyze the statistical significance of the results \cite{Conover1971}.}, CGAai and CGAbi perform similarly (p-value $=0.336$) and slightly better than the other classical alternatives (p-value $<0.001$), both obtaining $0.85$ average fidelity with a $0.02$ standard deviation, and UQCM performs better than these and other quantum alternatives (p-value $<0.001$), obtaining $0.91$ average fidelity with a $0.04$ standard deviation. Detailed analysis also shows that UQCM obtains better values than CGAs in all of the problem instances in terms of energy and in $90\%$ of the cases in terms of fidelity. 

\begin{figure}[tb]
\centering
\includegraphics[scale=1]{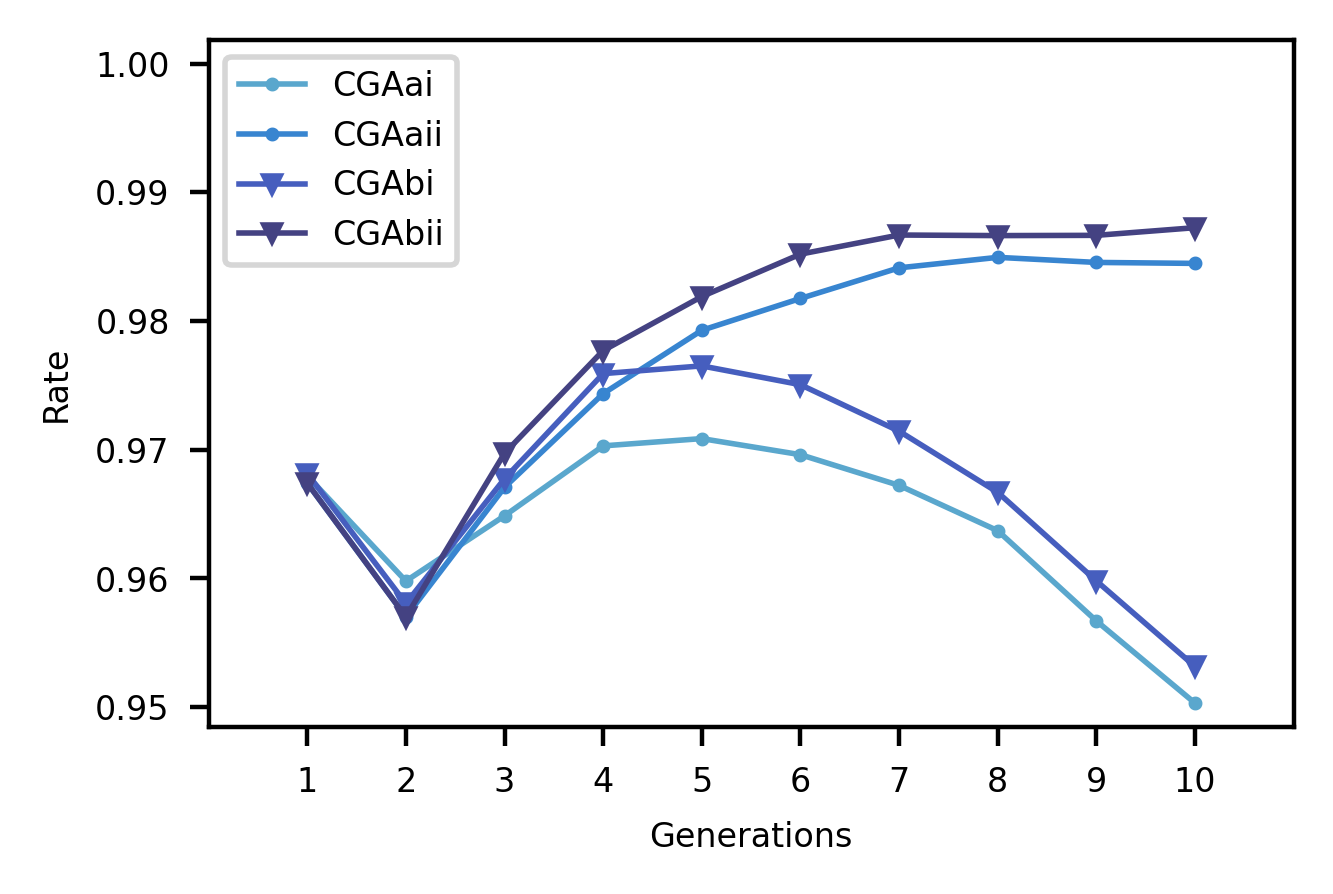}
\caption{Evolution of the ratio of initial populations that for each classical GA obtain lower or equal fidelity to the QGAunm in each generation step for the sample of 200 Hamiltonians. Visibly, UQCM obtains higher values than the classical variants with a ratio greater than $95\%$, although the advantage against CGAai and CGAbi decreases as the number of generations increases.}
\label{fig:Wintierate_UQCMnm_Stats}
\end{figure}

Figure~\ref{fig:Wintierate_UQCMnm_Stats} provides further insight into the comparison between the performance of the algorithms. We counted the rate of cases for each generation, varying with respect to the initial population, where the UQCM variant outperformed the others in terms of fidelity. The plot shows the values obtained by averaging out these rates for the sample of problem Hamiltonians. We observe that within $10$ generations QGAunm outperforms all the classical variants with a winning proportion above $0.95$. This supports the intuition that the exponential convergence of QGAunm provides an advantage in the early stages of the evolutionary process. However, the comparisons with respect to CGAai and CGAbi present a decrease in the rate of this quantum variant succeeding against them. In the following results, we explore the performance of the algorithms for longer evolutionary processes considering two particular Hamiltonians.

%

\subsubsection{Results for $H_{H_2}$ and $H_C$}

\begin{figure*}[!htb]
\centering
\includegraphics[scale=1]{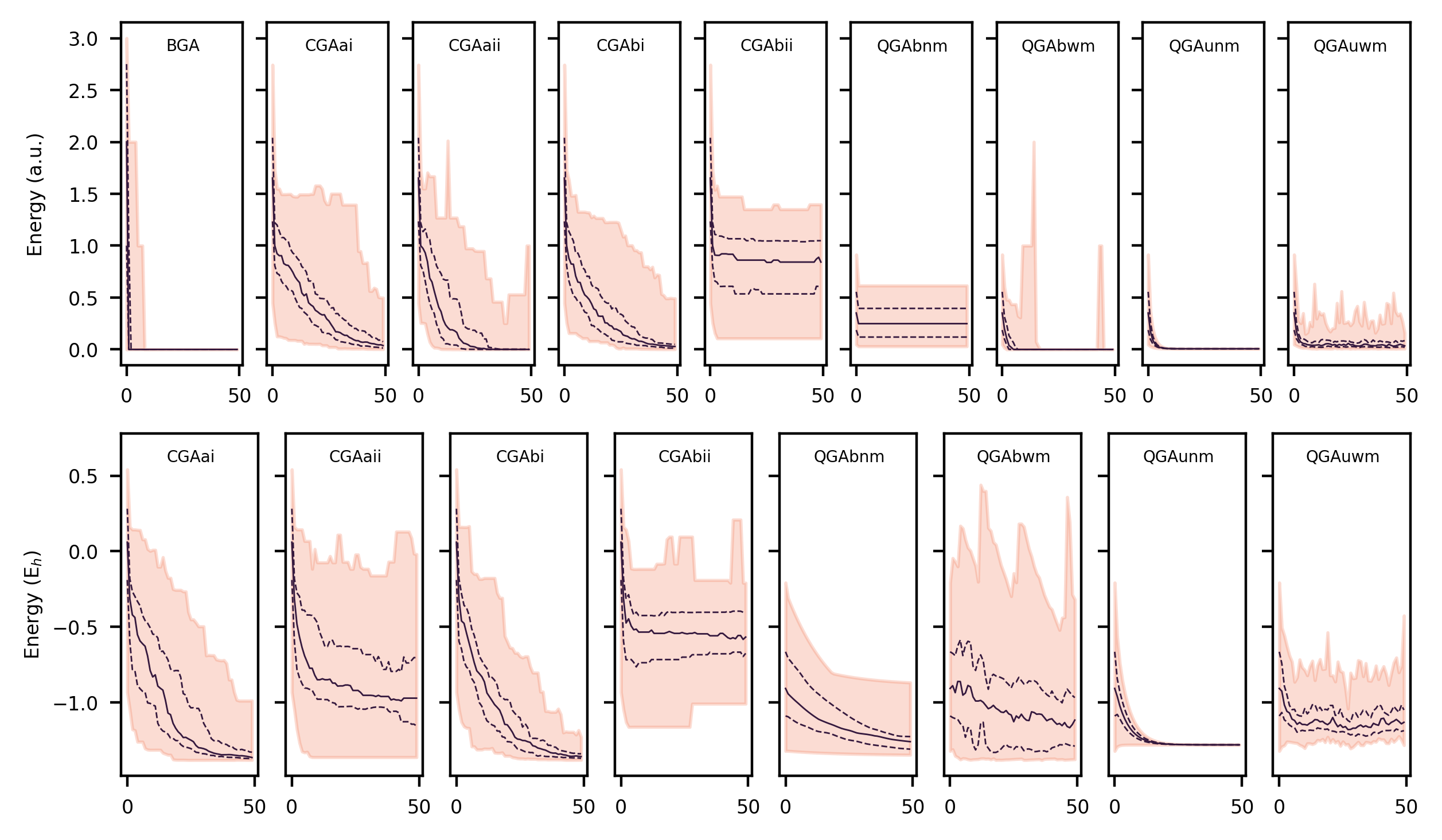}
\caption{
Evolution of the mean energy of the lowest-energy individual with respect to the diagonal Hamiltonian $H_C$ (first row) and the hydrogen molecule Hamiltonian $H_{H_2}$ (second row) for 50 generations ($x$-axis). All the GAs, except BGA, were initialized with the same 50 initial populations and evolved for 50 generations. The shaded area features the range of values found for the energy in each generation, the solid lines show the medians, and the dashed lines depict the $50\%$ confidence interval. The plot shows that BGA (only for $H_C$), QGAbwm, and QGAunm rapidly converge towards values near to the minimum, respectively $0$ and $-1.38$ for $H_C$ and $H_{H_2}$. CGAai, CGAaii, CGAbi, and QGAuwm seem to converge towards low energy states, though the convergence pattern is more erratic. In contrast, CGAbii and QGAbnm seem to reach an impasse in their evolution. These observations are consistent with the numeric values observed in Table~\ref{tab:qualitative_data} and the properties discussed in Section~\ref{sec:numerical_performance}.}
\label{fig:Hc_energy_comparison}
\label{fig:Hh2_energy_comparison}
\end{figure*}

For these particular cases, we have run the GAs with 50 generations to describe with greater detail the different convergence patterns of the QGA and the classical GAs. In this problem instances, the UQCM variant also has a single fixed point, whereas for the BCQO there is a convex set of fixed points. This is due to the spectral properties of the Hamiltonians $H_{H_2}$ and $H_C$ with respect to the cloning basis of the BCQO, which is also the reason BCQO produces better than average results.

\begin{table}
\centering
\caption{Average final fidelity and final energy for $H_C$ and $H_{H_2}$.}
\renewcommand{\arraystretch}{1.1}
\begin{tabular}{ccccc}
	& \multicolumn{2}{c}{$H_C$} & \multicolumn{2}{c}{$H_{H_2}$} \\[0.5ex]
\cmidrule(lr){2-3} \cmidrule(lr){4-5}
\rule{0pt}{2ex}
	   & Fidelity & Energy (a.u.)	& Fidelity & Energy (E$_h$) \\[1ex] \hline
\rule{0pt}{3ex}
BGA        & $1.00 (0.00)$ & $0.00 (0.00)$ &  -- & -- \\
CGAai      & $0.95 (0.07)$ & $0.07 (0.10)$ & $0.97 (0.04)$ & -$1.33 (0.07)$ \\ 
CGAaii     & $0.95 (0.16)$ & $0.05 (0.16)$ & $0.74 (0.18)$ & -$0.90 (0.33)$ \\ 
CGAbi      & $0.97 (0.07)$ & $0.04 (0.07)$ & $0.98 (0.02)$ & -$1.35 (0.03)$ \\ 
CGAbii     & $0.56 (0.16)$ & $0.81 (0.31)$ & $0.55 (0.11)$ & -$0.57 (0.20)$ \\ 
QGAbnm     & $0.80 (0.12)$ & $0.27 (0.16)$ & $0.92 (0.05)$ & -$1.25 (0.08)$ \\ 
QGAbwm     & $1.00 (0.00)$ & $0.00 (0.00)$ & $0.82 (0.15)$ & -$1.09 (0.24)$ \\ 
QGAunm     & $0.99 (0.00)$ & $0.01 (0.00)$ & $0.94 (0.00)$ & -$1.28 (0.00)$ \\ 
QGAuwm     & $0.95 (0.04)$ & $0.05 (0.04)$ & $0.83 (0.08)$ & -$1.10 (0.13)$ \\[1ex]  \hline
\end{tabular}
\label{tab:qualitative_data}
\end{table}

%

In Fig.~\ref{fig:Hc_energy_comparison} we represent the evolution of the energy for each GA. The shaded area depicts the range between the maximum and minimum values observed for different initial populations in each generation, the solid lines represent the medians, and the dashed lines the $50\%$ confidence interval. Additionally, in Table~\ref{tab:qualitative_data} we show the final average fidelity and average energy obtained by each algorithm.

Regarding the case of $H_C$, BGA performs best in terms of final result and convergence speed due to its reduced search space. This shows that if the eigenbasis of the Hamiltonian is known beforehand, introducing quantum degrees of freedom into classical strategies may not provide an advantage. For the $H_C$ problem instance, BCQO with mutation is identical to BGA except for the mutation subroutine, which also includes $Y$ and $Z$ gates, hence it presents similar results. For this reason, BCQO without mutation is identical to a BGA without mutation, thus lacking the ability to explore and rapidly stabilizes.

Regarding $H_{H_2}$, the algorithms CGAai, CGAbi, QGAbnm, and QGAunm perform similarly in terms of fidelity and energy of the final state, as presented in Table~\ref{tab:qualitative_data}. However, in terms of the convergence speed, Fig.~\ref{fig:Hh2_energy_comparison} suggests that QGAunm converges rapidly and holds the advantage observed above within $10$ generations.

\begin{figure}[tb]
\centering
\includegraphics[scale=1]{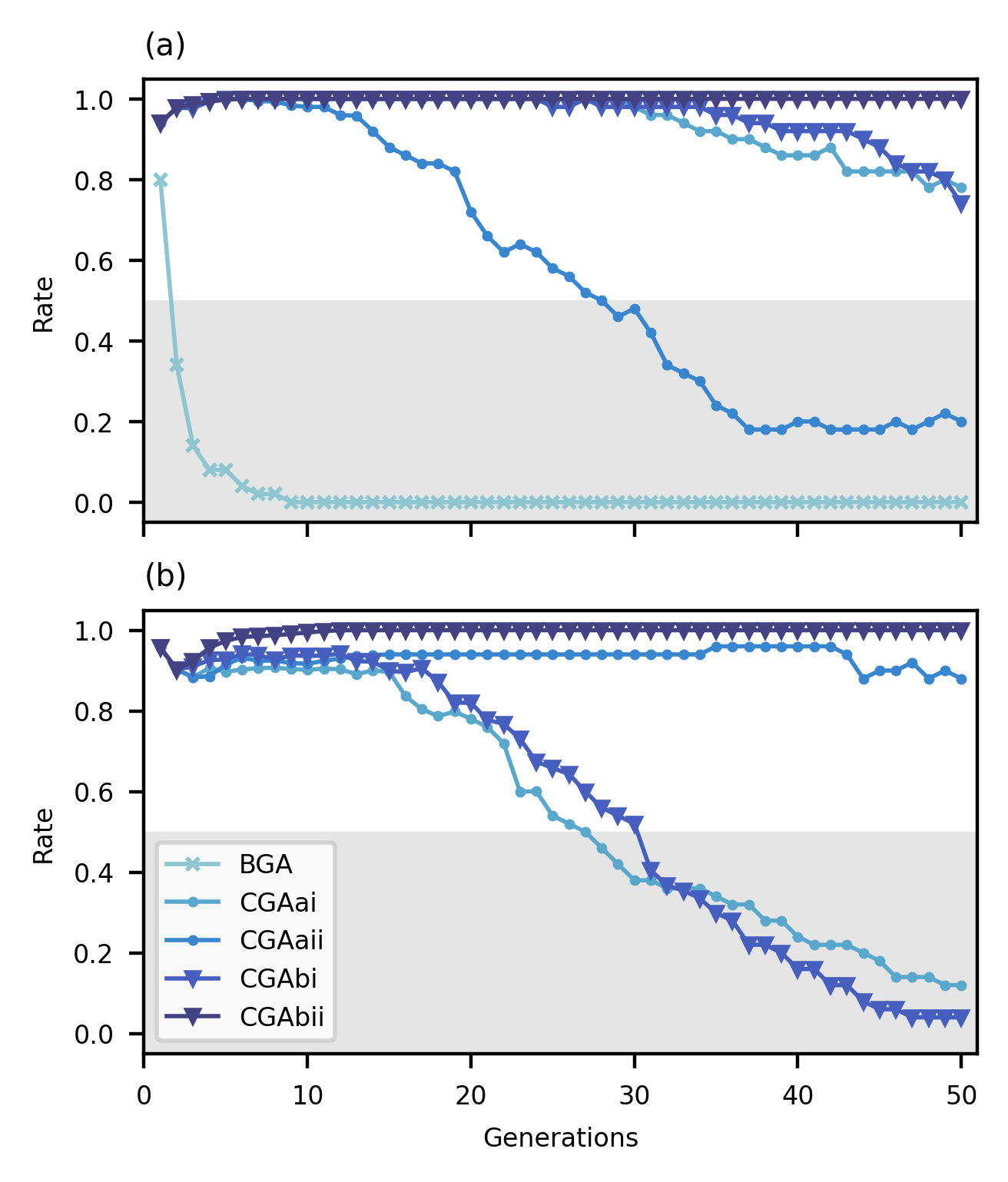}
\caption{Evolution of the ratio of initial populations that for each classical GA obtain lower or equal fidelity to the QGAunm in each generation step, (a) for the diagonal Hamiltonian $H_C$ and (b) for the hydrogen molecule Hamiltonian $H_{H_2}$. UQCM outperforms all the classical GAs, except BGA, until generation $20-30$, where it obtains better results than some CGAs less than half of the times.}
\label{fig:Wintierate_UQCMnm_Hc}
\label{fig:Wintierate_UQCMnm_Hh2}
\end{figure}

Fig.~\ref{fig:Wintierate_UQCMnm_Hh2} depicts the evolution of the rate of QGAunm obtaining better results than classical algorithms in terms of fidelity in each generation. We observe that it outperforms the classical variants in above $90\%$ of the cases within $10$ generations. However, the advantage decreases to $50\%$, the shaded area, within $20-30$ generations against CGAai and CGAbi for $H_{H_2}$, and CGAbi for $H_C$. Although in the case of $H_C$ BCQO with mutation performs better, this is due to the exceptional spectral properties of this Hamiltonian and it is not included in the figure.

\section{Conclusion}
\label{sec:conclusion}

We have provided a framework to numerically compare QGA variants and their nearest classical analogs. The size of quantum systems that we are able to simulate limits the regime of our comparison, which motivates the search for alternative benchmarks. Hence, our approach focuses on analyzing the effect of quantum resources, such as superposition, on the algorithm’s performance. We have studied quantum variants that differ in the employed quantum cloning machines and the inclusion or suppression of mutation. Whereas for the classical GAs, we considered two different encodings for the individuals, namely, bit-strings and complex vectors, and complex vectors were combined and mutated in two different manners.
Our comparison consisted in applying each of these GAs to different problem instances described by two-qubit Hamiltonians and analyzing the quality of their results in terms of their distance with respect to the optimal solution and their mean energy. Precisely, we have studied a sample of 200 randomly generated Hamiltonians complemented by two relevant cases, namely a diagonal Hamiltonian and the hydrogen molecule Hamiltonian. 
Our numerical analysis shows that for non-diagonal Hamiltonians the UQCM variant outperforms all the classical algorithms in terms of convergence speed, besides achieving near-optimal solutions. Indeed, within 10 generations the UQCM obtains higher fidelity than at least $90\%$ of the trials performed with other variants for the same Hamiltonians while obtaining an average fidelity of $0.91$ with a standard deviation of $0.04$. Of course, as UQCM converges towards a stationary state while classical variants actively persist in evolving, this advantage is reduced for a larger number of generations, which was around $20-30$ in the two relevant Hamiltonians studied.
In future work, developing the theoretical analysis of QGA variants could provide performance estimates for larger systems, and compare them against large-scale GA simulations. Remarkably, if this future analysis shows that the advantage in convergence speed also holds in larger search spaces, QGA efficiency would be verified for near-optimal optimization results.


\end{document}